# Possible phason-polaron effect on purely one dimensional charge order of Mo$_6$Se$_6$ nanowires

Xing Yang[1§], Jing-Jing Xian[1§], Gang Li[2], Naoto Nagaosa[3,4], Wen-Hao Zhang[1], Le Qin[1], Zhi-Mo Zhang[1], Jing-Tao Lü[1#], Ying-Shuang Fu[1*]

1. School of Physics and Wuhan National High Magnetic Field Center, Huazhong University of Science and Technology, Wuhan 430074, China

2. School of Physical Science and Technology, ShanghaiTech University, Shanghai 200031, China

3. RIKEN Center for Emergent Matter Science (CEMS), Wako, Saitama 351-0198, Japan

4. Department of Applied Physics, The University of Tokyo, Tokyo 113-8656, Japan

Email: [#] jtlu@hust.edu.cn,   [*] yfu@hust.edu.cn

[§] These authors contribute equally to this work.

**In one-dimensional (1D) metallic systems, the diverging electron susceptibility and electron-phonon coupling collaboratively drive the electrons into a charge density wave (CDW) state. However, strictly 1D system is unstable against perturbations, whose effect on CDW order requires clarification ideally with altered coupling to surroundings. Here, we fabricate such a system with nanowires of Mo$_6$Se$_6$ bundles, which are either attached to edges of monolayer MoSe$_2$ or isolated freely, by post-annealing the preformed MoSe$_2$. Using scanning tunneling microscopy (STM), we visualized charge modulations and CDW gaps with prominent coherent peaks in the edge-attached nanowires. Astonishingly, the CDW order becomes suppressed in the isolated nanowires, showing CDW correlation gaps without coherent peaks. The contrasting behavior, as revealed with theoretical modeling, is interpreted as the effect**

**of phason-polarons on the 1D CDW state. Our work elucidates a possibly unprecedented many body effect that may be generic to strictly 1D system but undermined in quasi-1D system.**

**INTRODUCTION**

When electrons are confined in an atomic chain, correlations among them get enhanced causing instability of the Fermi surface due to its perfect nesting and reduced screening[1]. This makes the systems susceptible to electron-electron and electron-phonon interactions, driving them into a rich variety of exotic correlated quantum states that are distinguished from 3D, such as the Tomonaga-Luttinger liquid, spin density waves, and CDW. In the CDW state, lattice symmetry is spontaneously broken via a Peierls distortion mechanism dictated by the electron-phonon coupling[2]. The electron density acquires a spatial modulation corresponding to twice the Fermi wave vector ($k_F$), and concomitantly opens a gap around the Fermi energy ($\varepsilon_F$). However, quantum and thermal fluctuations in strictly 1D system induce uncertainty in nuclear positions of the same order as that of the Peierls distortion[3], which tend to destroy the CDW coherence. Indeed, most experimental CDW orders are observed in quasi-1D systems[4-6], that are ensembles of chains contained either in a bulk 3D or a surface 2D form. Their inter-chain interaction is argued to stabilize the CDW state, but meanwhile causes complications. It is desirable to experimentally examine its existence in a truly 1D system. Although recent experiments have reported CDW orders in single silicide nanowires[7] and mirror twin boundary of monolayer $MoSe_2$[8] that are indeed electronically isolated 1D systems, they are still coupled to the lattice of supporting surroundings, whose phonons are essentially 2D. Thus, a 1D system with altered coupling to its surroundings is essential to uncover the embedded properties in inherent CDW order.

Creating such a system is challenging, because it should have 1D electronic character with reasonable regularity, defect-free from disorder and retain its 1D electronic property against altered coupling to the surroundings. Our strategy is to use transition metal monochalcogenide (TMM) nanowires, which have monolayer thickness, nanometer width and show metallic conductivity[9-12]. TMM is a polymorph of transition metal dichalcogenides (TMDs), which are widely studied due to their direct-band gaps with spin and valley polarization[13] at the single layer limit. Because of the polymorph, TMM nanowires can be obtained by judiciously tuning the stoichiometry and even achieving contacts to TMDs. However, TMM nanowires synthesized with previous methods suffer from either shape-irregularities or fixed coupling to surroundings[14-20].

Here, we present an approach for fabricating TMM $Mo_6Se_6$ nanowires by post-annealing the preformed TMD $MoSe_2$ monolayers prepared with molecular beam epitaxy[21] on a graphene-covered SiC(0001) substrate[22], producing straight nanowires with monolayer height and well-defined width. The nanowires are either attached to the $MoSe_2$ edges or isolated freely, allowing altered coupling to the environment. Equally importantly, the graphene substrate has a negligible interaction with the supported nanowires. Both conditions promise the system ideal for the study of purely 1D CDW order.

**RESULTS**

**Morphology and electronic structure of $Mo_6Se_6$ nanowires**

Figure 1A shows the morphology of MoSe$_2$ monolayer islands and a small fraction of bilayer islands all decorated with Mo$_6$Se$_6$ nanowires at the edges. Isolated nanowires bridging MoSe$_2$ monolayer islands are also seen. Statistics over 30 nanowires indicates their width is 3.35 ± 0.26 nm and uniform throughout individual ones. This reflects the nanowires are Mo$_6$Se$_6$ bundles containing mostly 3-wire or 4-wire in parallel. A zoom-in image of the bright edge displays the atomic resolution of the nanowire (Fig. 1B), showing a lattice constant of 0.44 nm along the nanowire. Its atomic resolution image is reproduced by our simulated STM image with density functional theory (DFT) based on the crystal structure of a 4-wire Mo$_6$Se$_6$ bundle (Fig. 1B, Supplementary Note 1). The atomic structure of such TMM wires prepared with a similar method has also been observed with a transmission electron microscope[19]. There is a Moiré period ~1.3 nm along the nanowires, implying their interface with MoSe$_2$ is atomically smooth, and a larger one ~5 nm that is likely from imperfect Moiré overlapping (Fig. S4A and S5).

Next, we show their tunneling spectra. While MoSe$_2$ has a band gap of ~2.2 eV [23], Mo$_6$Se$_6$ exhibits increased conductance with multiple peaks below $\varepsilon_F$, a relatively flat but finite conductance above $\varepsilon_F$, and an enhanced peak at ~1.6 eV with a splitting of ~0.2 eV (Fig. 1C, Fig. S6). The edge-attached and isolated nanowires exhibit similar spectra, indicating their electronic structure of 1D nature is well-conserved against the edge-attachment to MoSe$_2$ layer. Those spectroscopic features are captured by our DFT calculations of both a single Mo$_6$Se$_6$ wire (Fig. 1D) and a bundle containing 2-5 wires (3-wire bundle exemplified in Fig. 1E), demonstrating their bands are not substantially modified by the bundled structure. Comparison with the experiment reveals the nanowires are electron-doped ~0.3 eV by the substrate (Fig. 1C) resulting in a $k_F$ ~5.35 nm$^{-1}$ (Fig. 1E).

## CDW of Mo$_6$Se$_6$ nanowires

Intriguingly, spectra of a typical edge-attached nanowire (Fig. 2A) at 4.4 K reveal a salient gap spreading the entire nanowire (Fig. 2B, Fig. S4B). The gaps exhibit sharp coherence peaks with a gap size ($2\Delta$) of ~92 meV (Fig. 2C). Statistics over 17 nanowires indicate an average $2\Delta$ of 72.5 meV with a standard deviation of 29.3 meV, which is on a similar scale as that reported in the MoSe$_2$ mirror twin boundary[8]. The $2\Delta$ variation implies their different couplings to the MoSe$_2$ layers coming from the edge-attachment configurations. The observed gap cannot be stem from disorder-induced electron localization[24,25], because it has a U-shape with coherence peaks and a spatially identical size throughout each nanowire. To explore its CDW origin, we examine the real-space conductance of the nanowire around energies of the coherence peaks. There it exhibits periodic modulations of 0.64 ± 0.04 nm, which coincide with the calculated CDW period of $\pi/k_F$ (Fig. 1E) and are in anti-phase for the filled and empty states, respectively (Fig. 2D, another data set in Fig. S4, B to D). These features are consistent with the convention of CDW order. This ascription is augmented by elevating the temperature, where the CDW gap becomes intrinsically suppressed (Fig. 2E, Fig. S7, and Supplementary Note 2).

The CDW order in edge-attached nanowires has several properties. First, the CDW modulation is found incommensurate with the lattice and barely influenced by the strain-releasing defects (Fig. S4E). Second, there exists conductance inhomogeneity along the nanowire that correlates with the imperfect Moiré overlapping. The spatial scale is consistent with the CDW correlation length of $\hbar v_F/2\Delta \sim 2.5$ nm [26] (Fig. 2B, Fig. S4B). Third, a series of satellite peaks of $\sim 15 \pm 2$ meV spacing appear next to the coherence peaks (Fig. 2, C and D), which are ascribed to a phonon mode[8]. Fourth, the supporting graphene has negligible influence on the CDW order (Supplementary Note 3).

Interestingly, the isolated nanowire also exhibits a spectroscopic gap around $\varepsilon_F$ at 4.4 K (Fig. 3, A and B), which features enhanced conductance, related to an adjacent electronic state, below the lower gap edge, and power-law shaped onset of the conductance at the upper gap edge. While the gap shape shows some difference at each spectrum along the nanowire, they all have similar gap size ($\sim 130$ meV) except at both end-contacts, and surprisingly no coherence peaks (Fig. 3C). There is no charge modulation along the nanowire either (Fig. 3B). Similar spectroscopic gap has been observed in 10 isolated nanowires of different length. The gap is off-centered from $\varepsilon_F$ that may appear at either side (Fig. S10), and gets smeared intrinsically with increasing temperature (Fig. 3D). Since those characters are shared by the CDW gap of edge-attached nanowires, we are led to relate the spectroscopic gap of the isolated nanowire to CDW correlation. The vanishing oscillatory electron density with position itself in the presence of the well-defined gap strongly suggests the quantum disorder state due to the presence of phason. Therefore, it is reasonable to assume the incommensurate phason. The CDW correlated state without long-range coherent order[27] is analogous to the pseudogap in high temperature superconductors[28] and disordered superconductors[29], where the Cooper pairs are localized without forming long-range coherence.

## DISCUSSION

**Possible reasons for impacting the CDW coherence**

Several factors can impact the coherence of CDW order. Atomic disorder can destroy the phase coherence[30]. This, however, can be disregarded as the isolated nanowire has no defect-associated disorder. Finite length of the nanowires could affect the CDW coherence. This possibility is ruled out, because the isolated nanowires are connected to the edge-attached ones, which are long along the $MoSe_2$ edges. Quantum and thermal fluctuations of acoustic phonons and phason of the CDW, which correspond to atomic motions near the zero momentum and $2k_F$, respectively, can also destroy the CDW phase coherence[31,32]. Since thermal fluctuation is suppressed at 4 K, we focus on quantum fluctuation. Such zero-point fluctuation leads to a logarithmic singularity, changing the local spectral function $A(\omega)$ from the behavior $A(\omega) \propto (\omega - \Delta)^{-1/2}$ for the long-range ordered CDW to that of $A(\omega) \propto (\omega - \Delta)^{\alpha - 1/2}$ with the exponent $\alpha > 0$ (Supplementary Note 4). However, our estimated $\alpha \ll 1/2$ cannot change the diverging behavior to a finite one near the gap edge $\omega \sim \Delta$, due to the heavy atomic mass in the unit cell.

**Phason-polaron model on CDW of different dimensionality**

The above perturbative analysis gives weak influence of quantum fluctuations from acoustic phonons and phason on the CDW, which is inconsistent with the experimental observation. This suggests the quantum fluctuation from the above aspects is possibly strong, which leads us to consider a polaronic scenario. When an electron is injected from or to the STM tip, it may shake up the phonons or phasons, resulting in a polaronic effect (Fig. 4, A and B). We have estimated the polaron coupling constant to the acoustic phonon[33] and phason (Supplementary Note 4). By introducing phason fluctuations into the electron mean-field Hamiltonian, we arrive at an electron-phason coupling of similar form to acoustic phonons (Eq. S6), but the effective coupling is $\sim \frac{\varepsilon_F}{\Delta}$ fold enhanced. This is because the CDW introduces a rapid change of the electronic states within the narrow range $\xi^{-1} \sim \frac{\Delta}{\varepsilon_F} a^{-1}$ in the wavenumber space with $a$ being the lattice constant. (This $\xi$ is the coherence length of CDW.) Hence, the coupling to phason with wavenumber $q$ is proportional to $\xi q$ while that of the acoustic phonon to $aq$. Defining the dimensionless coupling constant $\gamma = \alpha_{phason}$, the spectral function of the phason-polaron is given by the same form $A(\omega) \propto (\omega - \Delta)^{\gamma - 1/2}$ simply replacing α by γ, but the physical origin is different. Our estimation of γ ~3 based on parameters from DFT calculations gives its upper bound and is consistent with the suppression of the diverging behavior of $A(\omega)$ at $\omega \sim \Delta$. This is in sharp contrast to the case of 2D phason-polaron, which gives $A(\omega) \propto e^{-\gamma}(\omega - \Delta)^{-1/2}$. Namely, the singularity remains with the reduced magnitude. With a cut-off phason energy $\hbar\omega_c = 10$ meV and $\gamma = 0.8$, the numerical CDW gaps in the presence of 2D and 1D phason-polaron including the thermal broadening effect (Fig. 4, C and D) nicely reproduce the experimental features considering the model's simplicity. This phason-polaron scenario can also predict the finite temperature effect, which in 2D case gives $A(\omega) \propto (\omega - \Delta)^{2\gamma(\frac{k_B T}{\hbar \omega_c}) - 1/2}$ with $k_B$ being the Boltzmann constant. Therefore, it is expected that the singularity of $A(\omega)$ disappears when $k_B T > k_B T_c = \hbar \omega_c / (4\gamma)$ giving $T_c \sim 36$ K. The temperature dependence of $A(\omega)$, which is quantified by the ratio of intensity between the coherence peak and the background, seems consistent with this expectation (Fig. 4E), giving another support for our picture. Note that our model is based on gapless phasons and assumes that the wires are infinitely long, which are not ideal in experiment. CDW pinning by disorder or commensurate lattice can introduce a phason gap[2]. Moreover, the finite length of the isolated wire introduces a crossover energy scale between the 1D and 2D phason. We have evaluated that their influence on phasons is negligible (Supplementary Note 5).

**Outlook**

Our study envisions several future studies. The phason-polaron effect suggested here should be generic to strictly 1D CDW states, which can be examined in more experimental systems. $Mo_6Se_6$ bundles with increased width, on the other hand, are expected to induce a dimensional crossover from strictly 1D to quasi-1D, whose influence on the CDW states can be studied systematically. Moreover, the polaronic effect with different dimensionality may also interplay with other correlated states in 1D system, such as spin density waves, Tomonaga-Luttinger liquid[34], etc., which opens up new directions for in-depth investigations.

**MATERIALS AND METHODS**

**Sample preparation.**

SiC(0001) substrate: The SiC substrate (MTI corporation) was firstly degassed at 900 K for at least 3 hours to remove organic adsorbates. Then, it is flashed to 1220 K for 2 min while facing a Si source heated at 1470 K. Five cycles of similar flashing procedures is performed. Subsequently, the Si source is turned off. And the SiC is flashed to a higher temperature of 1670 K for 5 min to desorb Si atoms and form a graphene layer.

$MoSe_2$ thin films: The $MoSe_2$ films are grown by co-depositing high purity Se (purity 99.999%) and Mo (purity 99.95%) atoms from a Knudsen-cell and an electron-beam evaporator respectively, while the graphene-covered SiC substrate is kept at 530 K for a duration of 10 min. After that, the sample is annealed at 870 K for 10 min to crystalize the $MoSe_2$ films. The flux ratio of Se and Mo is about 10:1, and the excess Se atoms are desorbed from the substrate during the growth.

$Mo_6Se_6$ nanowires: To form $Mo_6Se_6$ nanowires, the sample is further heated to 870 K *in situ*. During this process, the Se atoms at $MoSe_2$ edges preferentially evaporate upon the annealing treatment, changing the stoichiometry of Mo and Se at the edges and driving a transition to the $Mo_6Se_6$ phase. Since the reaction occurs at the defined location, the straight edges of $MoSe_2$ layers provide excellent templates for the growth of $Mo_6Se_6$ nanowires. By tuning the annealing time, we can control the ratio of $Mo_6Se_6$ terminated edges (Fig. S3, A to C). A minimum of 8 hours is needed to fully saturate the $MoSe_2$ edges with $Mo_6Se_6$ nanowires. Further extended annealing degrades partial $MoSe_2$ layers, and meanwhile forms some isolated $Mo_6Se_6$ nanowires connecting to $MoSe_2$ islands at two ends (Fig. S3D). These wires are tantalizing for making interconnects to the monolayer circuits.

**STM measurement.**

The experiments were performed with a cryogenic custom-made Unisoku STM[35]. An electro-chemically etched W wire was used as the STM tip. Prior to measurements, the tip was characterized on a Ag(111) multi-layer films grown on a Si(111) substrate, which has been cleaned by several flashing cycles to 1500 K. The tunneling spectra were obtained by lock-in detection of the tunneling current with a modulation voltage at 983 Hz feeding into the sample bias.

**DFT calculation.**

The electronic structure and the DOS of $Mo_6Se_6$ nanowires were calculated in the framework of DFT within the generalized gradient approximation[36] and local density approximation, where both approximations deliver identical results. The projector-augmented-wave method implemented in the Vienna Ab Initio Simulation Package was employed with an energy cutoff of 500 eV[37,38]. A supercell with vacuum spaces of 20 Å along x- and y-direction is employed with a $1 \times 1 \times 9$ $k$-point mesh. Phonopy[39] package is further used to calculate the phonon band structure using a $1 \times 1 \times 8$ supercell and $1 \times 1 \times 2$ $k$-point mesh. The calculations are performed at the Shanghai Supercomputing Center, the HPC Platform of ShanghaiTech University Library and Information Services and School of Physical Science and Technology.

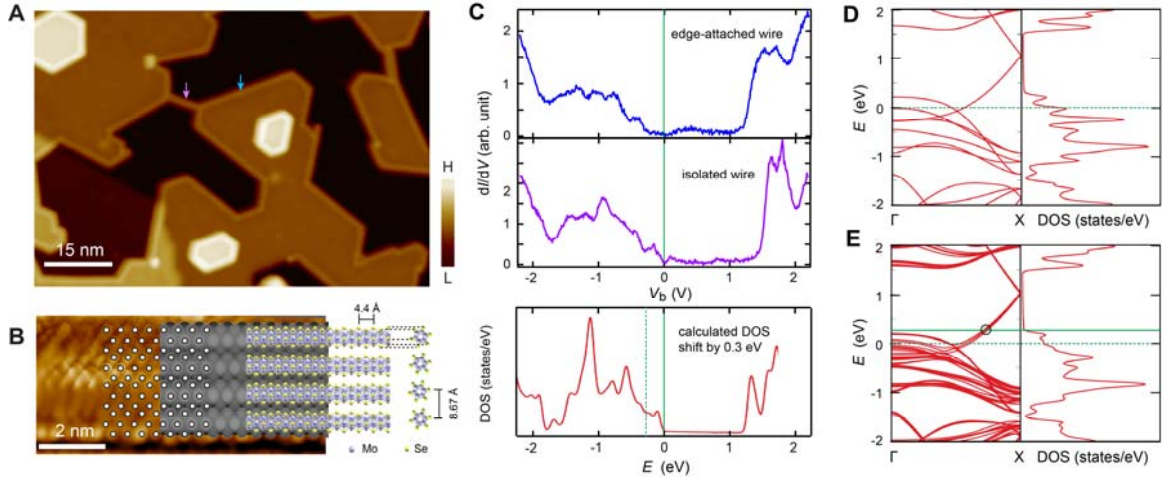

**Fig. 1. Morphology and electronic structure of $Mo_6Se_6$ nanowires.** (**A**) STM topography ($V_s$ = 2.2 V, $I_t$ = 10 pA) of $Mo_6Se_6$ nanowires. An exemplified isolated (edge-attached to $MoSe_2$ monolayer) nanowire is indicated with a purple (cyan) arrow. (**B**) Atomic resolution image ($V_s$ = 150 mV, $I_t$ = 100 pA) of an edge-attached nanowire, which is superimposed with the crystal structure of a 4-wire bundle (top view (left) and side view (right)) and its DFT-simulated STM image. The white dots mark selected positions of the imaged atoms. (**C**) Typical tunneling spectra of an edge-attached (blue curve) and an isolated (purple curve) nanowire of 4-wire bundle ($V_s$ = 2.2 V, $I_t$ = 100 pA, $V_{mod}$ = 14.14 mV$_{rms}$, $T$ = 77 K). (**D** and **E**) DFT-calculated band structure and density of states (DOS) of a single $Mo_6Se_6$ wire (D) and a 3-wire bundle (E), whose crystal structure is same as that in (B). The solid and dashed green lines in (C to E) mark $\varepsilon_F$ of the measured and calculated nanowires, respectively. The black circle marks $k_F$. The DOS of (E) is shown in (C) for comparison.

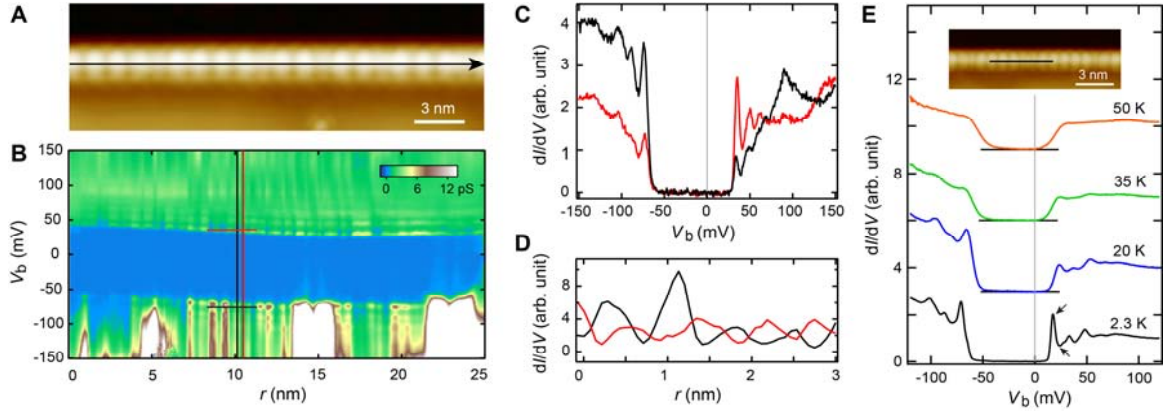

**Fig. 2. CDW state of edge-attached Mo$_6$Se$_6$ nanowire**. (**A**) STM image ($V_s$ = 600 mV, $I_t$ = 10 pA) of an edge-attached Mo$_6$Se$_6$ nanowire (4-wire bundle). (**B**) 2D conductance plot taken along the black line in (A) ($V_s$ = 150 mV, $I_t$ = 100 pA, $V_{mod}$ = 2.12 mV$_{rms}$., $T$ = 4.4 K). (**C**) Two typical spectra showing CDW gap. The black (red) curve is extracted from (B) along the vertical black (red) line, corresponding to two adjacent conductance extrema. (**D**) Line profiles of the conductance plot in (B) along the horizontal red and black line, respectively. (**E**) Tunneling spectra of an edge-attached nanowire (3-wire bundle) at different temperatures ($V_s$ = 150 mV, $I_t$ = 100 pA, $V_{mod}$ = 1.41 mV$_{rms}$), whose image ($V_s$ = 500 mV, $I_t$ = 10 pA) is shown in inset. Each spectrum is an average of the spectra along the black line (5 nm long) of the nanowire. The spectra have been offset vertically for clarity. Zero conductance for each spectrum is marked with a black line.

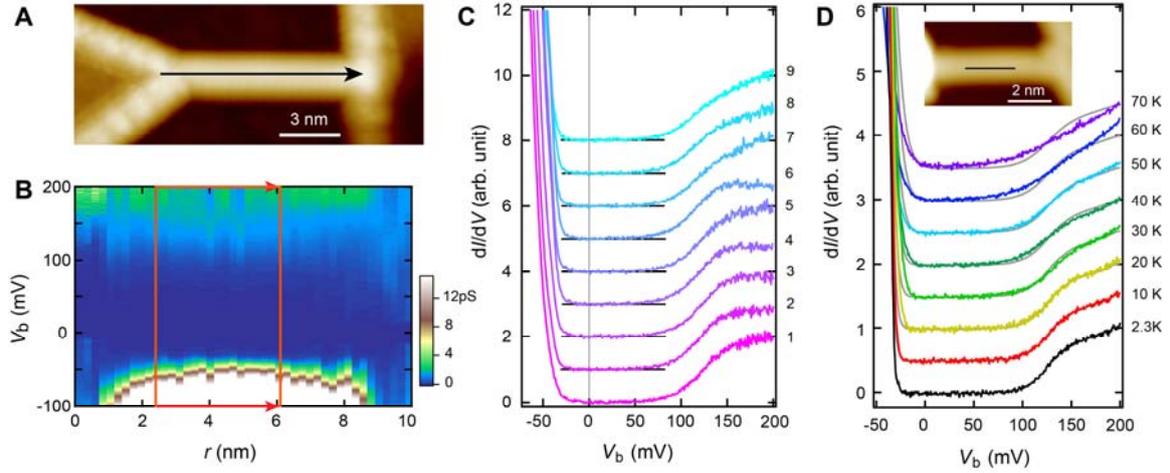

**Fig. 3. CDW correlation in isolated Mo$_6$Se$_6$ nanowire.** (**A**) STM topography ($V_s$ = 0.5 V, $I_t$ = 20 pA) of an isolated Mo$_6$Se$_6$ nanowire (3-wire bundle). (**B**) 2D conductance plot taken along the black line in (A) ($V_s$ = 200 mV, $I_t$ = 100 pA, $V_{mod}$ = 1.41 mV$_{rms}$, $T$ = 4.4 K). (**C**) Tunneling spectra extracted from the red rectangle of (B) in an equally spaced manner. The black lines mark the zero conductance for each spectrum. (**D**) Tunneling spectra of an isolated Mo$_6$Se$_6$ nanowire (3-wire bundle), whose image ($V_s$ = 2.2 V, $I_t$ = 10 pA) is shown in inset, at different temperatures ($V_s$ = 200 mV, $I_t$ = 50 pA, $V_{mod}$ = 1.41 mV$_{rms}$). Each spectrum is an average of the spectra taken along the black line (2.3 nm long) of inset. For comparison, the spectra are superimposed with the simulated spectra (grey curves) of 2.3 K with thermal broadening effect at each temperature. The spectra in (C) and (D) have been offset vertically for clarity.

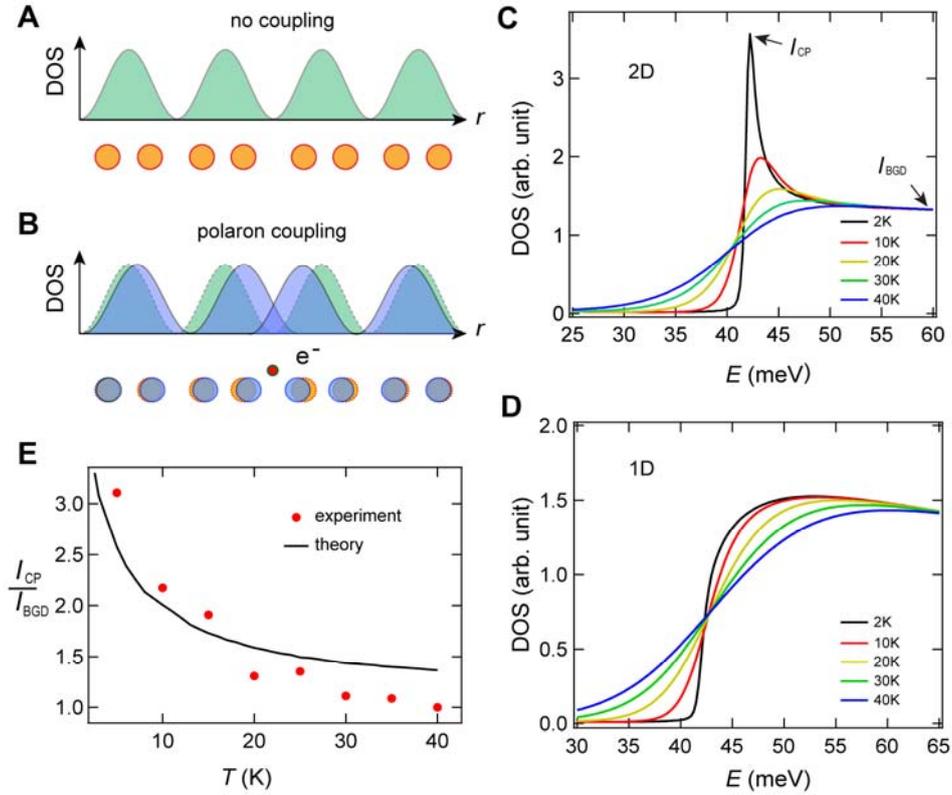

**Fig. 4. Phason-polaron effect on 1D CDW state.** (**A**) Schematics showing the CDW wave (green shaded curve) and a chain of atoms with Peierls distortion (orange balls). (**B**) Schematics showing formation of phason-polaron. The red ball represents an additional charge. The blue (orange) balls illustrate the chain of atoms after (before) the phason-polaron formation. The resulting phason is depicted with blue shaded curve. (**C and D**), Simulated DOS showing spectra of 2D (C) and 1D phason-polarons (D) after considering thermal broadening, corresponding the cases of isolated and edge-attached nanowires, respectively. For calculating the phason-polaron CDW spectra, we choose $\gamma = 0.8$ and $\hbar\omega_c = 10$ meV, $\Delta = 42$ meV, and damping energy= 0.1 meV. Only empty states of the spectra are shown for clarity. (**E**) Comparison between the experimental and theoretical ratio of intensity between the coherence peak ($I_{CP}$) and the background ($I_{BGD}$) (black arrows in C and Fig. 2E) in the 2D case.

**Acknowledgement:** The authors thank T. Hanaguri, H.W. Liu, X. Liu, and D.L. Feng for discussions. **Funding:** This work is funded by the National Key Research and Development Program of China (Grant No. 2017YFA0403501, 2016YFA0401003), the National Science Foundation of China (Grants No. 11874161, No. 11522431, No. 11474112, No. 11774105, No. 21873033), and Program for Professor of Special Appointment (Shanghai Eastern Scholar) ), and JST CREST Grant Number JPMJCR1874 and JPMJCR16F1, and JSPS KAKENHI Grant numbers 18H03676 and 26103006, Japan.. **Author contributions**: X.Y., J.J.X., L.Q., Z.M.Z., W.H.Z. carried out the experiments. G.L. and J.T.L. did the DFT calculations. N.N. and J.T.L. constructed the theoretical modeling. Y.S.F., X.Y., J.J.X., J.T.L., and N.N. analyzed the data. Y.S.F., N.N. wrote the manuscript with contributions from J.T.L. and G.L. All authors commented the manuscript. **Competing interests:** The authors declare that they have no competing interests. **Data and materials availability:** All data needed to evaluate the conclusions in the paper are present in the paper and/or the Supplementary Materials. Additional data related to this paper may be requested from the authors. Correspondence and requests for materials should be addressed to Y.S.F. and J.T.L..